**Characterization of the infectious reservoir of malaria with an agent-based model calibrated to age-stratified parasite densities and infectiousness**

Jaline Gerardin[1]*, André Lin Ouédraogo[1,2], Kevin A. McCarthy[1], Bocar Kouyaté[2,3], Philip A. Eckhoff[1], Edward A. Wenger[1]

[1]Institute for Disease Modeling, Intellectual Ventures, 1555 132nd Ave NE, Bellevue WA 98005 USA

[2]Department of Biomedical Sciences, Centre National de Recherche et de Formation sur le Paludisme, Ouagadougou, Burkina Faso

[3]Ministère de la Santé, Ouagadougou, Burkina Faso

*To whom correspondence should be addressed. Email: jgerardin@intven.com

**Abstract**

**Background** Elimination of malaria can only be achieved through removal of all vectors or complete depletion of the infectious reservoir in humans. Mechanistic models can be built to synthesize diverse observations from the field collected under a variety of conditions and subsequently used to query the infectious reservoir in great detail.

**Methods** The EMOD model of malaria transmission was calibrated to prevalence, incidence, asexual parasite density, gametocyte density, infection duration, and infectiousness data from 9 study sites. The infectious reservoir was characterized by diagnostic detection limit and age group over a range of transmission intensities with and without case management and vector control. Mass screen-and-treat drug campaigns were tested for likelihood of achieving elimination.

**Results** The composition of the infectious reservoir by diagnostic threshold is similar over a range of transmission intensities, and higher intensity settings are biased toward infections in children. Recent ramp-ups in case management and use of insecticide-treated bednets reduce the infectious reservoir and shift the composition toward submicroscopic infections. Mass campaigns with antimalarial drugs are highly effective at interrupting transmission if deployed shortly after ITN campaigns.

**Conclusions** Low density infections comprise a substantial portion of the infectious reservoir. Proper timing of vector control, seasonal variation in transmission intensity, and mass drug campaigns allows lingering population immunity to help drive a region toward elimination.

**Keywords**

mechanistic model, infectious reservoir, diagnostics, infectivity, drug campaigns

**Background**

Malaria is a global disease responsible for hundreds of thousands of deaths each year[1]. In the last decade, many regions have made considerable progress in malaria control and are now working toward local elimination of malaria[2, 3]. Technical strategies for elimination differ from those for control, and it is critical to understand the factors that lead to successful outcomes and make the best use of available resources.

Elimination of malaria requires interrupting transmission. Complete depletion of the infectious reservoir of malaria parasites requires clearing all malaria infections in the human population. In regions where malaria is endemic, accumulated exposure to infection leads individuals to develop immunity to clinical symptoms. Proper identification of parasite carriers is therefore confounded by a population of asymptomatic people.

Current rapid diagnostic tests (RDT) can quickly and cheaply identify individuals with parasite densities above 50-200 parasites per µL[4]. However, in endemic areas, a significant portion of the population harbors infections that are undetectable by RDT, and microscopy or molecular methods are required to identify infected individuals. Although capable of detecting infections as low as 0.01-0.05 parasites per µL, even the most sensitive molecular tests today cannot detect all infections[5].

Asexual parasite density, gametocyte density, and human infectiousness are known to be linked[6, 7]. Higher asexual densities lead to higher gametocyte densities, which increase the chance of infecting mosquitoes. However, the relationship between gametocyte density and infectiousness to mosquitoes is complex, as individuals with high gametocyte density may infect few or no mosquitoes in xenodiagnostic studies, and individuals with no observable gametocytes have been observed to be infectious[8, 9]. Human sexual-stage and transmission-blocking immunity have been proposed to play roles in moderating the relationship between gametocyte density and infectiousness[10], but the magnitude and duration of sexual stage immunity effects remain unclear.

While most malaria infections in areas of even moderate transmission are asymptomatic[11], the relative contribution to the infectious reservoir of symptomatic and asymptomatic, patent and subpatent, and adult and child populations remains an area of active investigation [12, 13]. Subpatent individuals, even if they outnumber patent individuals, are less infectious to mosquitoes. In an elimination setting, it is critical to understand how success may or may not depend on appropriate targeting of subpatent infections, as sensitive diagnostics can be very resource-intensive.

Mathematical modeling of malaria transmission and immunity can begin to answer some of these questions[12, 13]. A single mechanistic model can tie together apparently contradictory field data that describe malaria transmission in a variety of settings and demographic groups. With a detailed model of within-host infection, including parasite and gametocyte life cycles and densities as well as acquired immunity, the relative contribution to the infectious reservoir due to each subset of the population can be compared and implications for control and elimination can be quantitatively assessed.

A recent longitudinal study in a high-transmission area of Burkina Faso used molecular methods to quantify asexual and gametocyte densities and membrane feeding tests to measure infectiousness to mosquitoes, providing invaluable data for model calibration[14]. Using this dataset and an agent-based stochastic model of malaria transmission, we first calibrate adaptive immunity parameters to best capture the age-dependent distribution of asexual parasite and gametocyte densities. Next, we use the membrane feeding data to calibrate model parameters governing human infectiousness to mosquitoes. The calibrated model is then used to characterize the composition of the infectious reservoir by detection threshold and age group through seasonal variations and at a range of transmission intensities. We describe how the infectious reservoir changes when case management and vector control are deployed. Finally, we discuss the implications for elimination campaigns by testing how mass drug campaigns may successfully achieve elimination by targeting all or only a portion of the infectious reservoir.

**Methods**

*Malaria transmission model*

All simulations were conducted with EMOD DTK v1.6, an agent-based mechanistic model of malaria transmission. During the blood stage of malaria infection, asexual parasite density triggers innate and adaptive immunity within the host. Innate immunity stimulates cytokine production, leading to fever in a density-dependent manner and limiting the maximum parasite density. Adaptive immunity is modeled using three types of antigens: *P. falciparum* erythrocyte membrane protein 1 (PfEMP1) variants, merozoite surface proteins (MSP), and minor epitope variants. Fuller discussion of immune system modeling within the EMOD model is described in [15-17]. Asexual parasite and gametocyte densities are tracked within each host. Gametocytes differentiate from infected red blood cells and mature in five stages over ten days, and a fraction of gametocytes are lost at each stage of maturation from implicit immune effects. Mature stage gametocytes are taken up in a mosquito blood meal with density-dependent probability, and human and mosquito immune factors implicitly limit gametocyte survival within the mosquito. Infectiousness of an individual human is calculated as a xenodiagnostic experiment measuring the fraction of mosquitoes infected after feeding on the individual. The EMOD model is available for download at http://idmod.org/software.

*Calibration of asexual parasite and gametocyte densities by age and season*

Calibration of parasite densities built on previous work in [18]. Data from nine study sites were used, including four sites with age-stratified prevalence data—Namawala in Tanzania and Matsari, Rafin Marke, and Sugungum in Nigeria—and two sites with age-stratified clinical incidence data—Dielmo and Ndiop in Senegal—used in previous work [19-22]. Recently available data from two sites, Dapelogo and Laye in Burkina Faso, provided age- and season-stratified asexual parasite and gametocyte densities as measured by molecular methods [14]. Malariatherapy data on peak parasitemia and gametocytemia as well as duration of parasitemia and gametocytemia for single infections in naïve adults were also included in the calibration[23].

Parameters under calibration included the maximum number of simultaneous infections, number of PfEMP1 variants available to the parasite population, switching rate between PfEMP1 variants, number of MSP variants, number of minor epitope variants, killing strengths of anti-MSP and anti-minor epitope immune responses, production rate of gametocytes, and survival rate of gametocytes as they progress through stages of maturation.

Prevalence and incidence sites were simulated and likelihoods of calibration parameters were calculated as described in [18] All sites other than malariatherapy were simulated with forced entomological inoculation rate (EIR); no vectors are simulated, and instead the population experiences a pre-set number of infectious bites per month. Monthly EIR for Dapelogo and Laye followed Burkina Faso seasonality and were based on available entomological data, with an annual EIR of 300 for Dapelogo and 30 for Laye [24]. Malariatherapy simulations included a single infectious bite on day 0 with no other infectious bites.

In the likelihood function used to calibrate asexual parasite and gametocyte densities, data from each age group (under 5, 5-15, over 15), season (start of wet season, peak of wet season, dry season), study site (Dapelogo, Laye), and parasite stage (asexual, gametocyte) were considered separately (**Figure 1**). Asexual parasite and gametocyte densities were collected in simulations on July 1 (start of wet season), September 1 (peak of wet season), and March 1 (dry season), binned for each age group, and compared with field data with a Dirichlet-multinomial distribution.

For each parameter set $\vec{\theta}$, the parasite/gametocyte density distributions in each age group, site, and season are binned into 6 density bins. The likelihood of parameter set $\vec{\theta}$ given the data $d$ (consisting of $n_d$ total measurements, $\vec{k_d}$ counts in each bin, $\sum_{i=1}^{6} k_{d,i} = n_d$) is approximated as follows. The prior distribution for the multinomial probability vector $\vec{p}$ associated with the 6 bins is initially assumed to be a symmetric Dirichlet distribution with concentration parameter 1. The simulated distribution of counts $\vec{k_s}$ is used to inform the posterior distribution of $\vec{p}$:

$$Dir(\vec{p}|\vec{1}) \rightarrow Dir(\vec{p}|\vec{1} + \vec{k_s})$$

The likelihood of $\vec{\theta}$ is then approximated using the posterior predictive distribution computed by marginalizing over $\vec{p}$:

$$\mathcal{L}(\vec{\theta}|d) = P(d|\vec{\theta}) = \int P(d|\vec{p})P(\vec{p}|\vec{\theta})d\vec{p} = \int Mult(\vec{k_d}|n_d,\vec{p})Dir(\vec{p}|\vec{1} + \vec{k_s})d\vec{p} = DirMult(\vec{k_d}|n_d,\vec{1} + \vec{k_s})$$

The joint likelihood is the product of likelihoods for each age group, season, study site, and parasite stage, which is also multiplied with the likelihoods from prevalence, incidence, and malariatherapy comparisons to determine the overall likelihood of each parameter set. Likelihoods for prevalence and incidence are approximated using a methodology similar to that described above (detailed in [18]), using beta-binomial and gamma-Poisson distributions, respectively.

For calibration to malariatherapy data, patients who received antimalarial treatment were removed from the comparison data. All patients were assumed to have only a single infection. Distributions of peak parasitemias, peak gametocytemias, parasitemia duration, and gametocytemia durations were compared to simulation using a Dirichlet-multinomial likelihood as described above.

Parameter sampling was done by incremental mixture importance sampling (IMIS), as described in [25]. Final best-fit parameter values are shown in **Table S1 in Additional file 1** and comparison with study site data shown in **Figure S1 in Additional file 2**. Malariatherapy data was weighted at 10% relative to data from endemic areas in the likelihood function, as full weighting of malariatherapy data resulted in poorer fitting to prevalence data, particularly at the Rafin Marke study site (**Figure S2 in Additional file 2**). Infections from malariatherapy studies appear to have relatively short durations compared to what would be expected based on data collected from endemic sites. The malariatherapy dataset used in calibration, which did not include any patients who received curative treatment with drugs, may be biased toward milder infections. Strain differences between parasites used for malariatherapy and parasites in endemic regions may also account for the discrepancies. Differences in infection duration between strains has been previously noted, with El Limon showing long infections, McLendon showing short infections, and Santee Cooper showing both long and short infections[17].

*Calibration of infectiousness by age and season*

Human infectiousness to mosquitoes was calibrated to membrane feeding data stratified by age, season, and study site [14]. Simulations used the immunity, infection, and gametocyte development parameters in **Table S1 in Additional file 1**. A population of 1000 individuals of all ages was subjected to 50 years of forced EIR to initialize age-appropriate immunity, with EIR values as described above for the Dapelogo and Laye sites. Calibration also used forced EIR and no vectors.

Two parameters, the survival rate of gametocytes in the mosquito and the maximum probability a mosquito becomes infected, were calibrated using IMIS as described above. For each sampled parameter set, the distribution of fraction mosquitoes infected was compared to field data with a Dirichlet-multinomial as described above. Infectiousness by age group (under 5, 5-15, over 15) and season (start of wet season, peak of wet season, dry season) were considered separately, and likelihoods were multiplied. Calibrated parameter values are shown in **Table S2 in Additional file 1** and simulated infectiousness compared with field data in **Figure 2**.

*Characterization of the infectious reservoir in full vector transmission model*

Simulations of the infectious reservoir were conducted with a full vector transmission model in a Burkina Faso climate with vector abundance calibrated to achieve annual EIR between 200 and 300. Vector species included *A. gambiae* and a small amount of *A. funestus*, and larval habitats were modeled as in [26] with enough constantly available habitat to ensure vector survival through the dry season. Vector feeding behavior was 95% anthropophagic and 95% endophagic; other vector life cycle parameters were modeled as in [26]. Population demographics followed the Burkina Faso age structure [27], and all simulations were performed on a well-mixed population of 1000 individuals. Larval habitat was scaled to simulate settings with lower transmission intensity (**Table S3 in Additional file 1**). Simulation of each transmission intensity was repeated for 1000 stochastic realizations, and all figures show mean measurement values. Populations were simulated for 50 years at each transmission intensity in the absence of interventions to initialize age-appropriate immunity. Importation of cases occurred at a rate of 1/year. All simulations included 1 year of equilibration prior to 3 years of measurement.

Individuals were assigned to detection groups on each day of the detection period. Each individual was tested for asexual parasite positivity at 3 diagnostic sensitivity levels: 100, 10, and 0.05 asexual parasites per µL and assigned to the least sensitive diagnostic to which they tested positive. Individuals tested positive if a random draw from a Poisson distribution centered at the true number of asexual parasites in 1/(diagnostic sensitivity) µL of blood was nonzero. Individuals testing negative by all diagnostics, but nonetheless infected with asexual parasites, were assigned to the undetectable group.

Host size is known to predict mosquito biting behavior, and an age-dependent biting risk was included to approximate surface area dependence[28-30]. Composition of the infectious reservoir is calculated by normalizing the scaled human infectiousness. Uncertainty in parameter values and stochastic variations between simulation runs do not qualitatively change the composition of the infectious reservoir (**Figure S3 in Additional file 2**).

*Modeling case management and campaigns with insecticide-treated bednets (ITNs)*

Several simulation scenarios are modeled: baseline "no intervention" scenarios with no case management or ITNs; case management scenarios where symptomatic individuals could access treatment with antimalarial drugs, but no ITNs; ITN campaign scenarios where ITNs were distributed to the population in an age-dependent manner, but no case management was available; and scenarios that included both ITN distribution and case management.

Case management was available to simulated individuals with clinical or severe malaria. A clinical case occurred when an individual's body temperature increased by at least 1.5°C, and severe cases occurred when fever, anemia, or parasite density exceeded thresholds described in [18]. Ninety-five percent of the population had any access to case management. Of the population with access to case management, individuals with clinical malaria had 60% probability of seeking care for a given episode. Individuals with severe malaria had 95% probability of seeking care. All individuals seeking care received treatment with artemether-lumefantrine (AL) within 3 days for clinical cases or 2 days for severe cases, and all treated individuals completed the full course of treatment with full adherence. Age-dependent dosing, pharmacokinetics, and pharmacodynamics of AL were modeled as described in[31]. Simulations with case management allowed case management during both the equilibration and measurement periods.

ITNs were distributed at birth with 90% coverage and were also mass distributed on day 200 of the first measurement year with 80% coverage for children under 10 and 50% for individuals over 10 years of age. ITNs had an initial blocking rate of 80% and killing rate of 60%, with efficacy decaying exponentially with half-lives of 2 years and 4 years respectively. When ITNs were present, the contribution of ITN-protected individuals to the infectious reservoir was reduced by the ITN blocking rate after scaling by surface area as discussed above.

Annual EIR was measured by summing the daily infectious bites per person over the second measurement year. For the same amount of larval habitat, case management and ITNs reduced the apparent EIR relative to simulations without interventions. When case management and/or ITNs were present, the plotted EIR is the apparent EIR experienced by the simulated population.

*Simulation of mass drug campaigns*

Mass drug campaigns were conducted with dihydroartemisinin-piperaquine (DP), with improved pediatric dosing as described in [32] and pharmacokinetics and pharmacodynamics as in [31]. All individuals receiving DP adhered to the full treatment course. Mass campaigns were conducted during the second and third measurement years in 3 rounds per year separated by 6 weeks, with the first round occurring on day 60 of the second measurement year. All covered individuals received DP on the same day. Coverage was independent between rounds. Importation of cases occurred at a rate of 1/year.

Mass campaigns were tested at apparent EIRs between 0.07 and 40 for settings with no other interventions and settings with case management and ITN use as described above. Drug campaign coverage was tested at 0%, 20%, 40%, 60%, 80%, and 100%. Interpolations were calculated with the SciPy v0.14.1 interpolate function in Python 2.7. Mass screen-and-treat (MSAT) campaigns used diagnostics with sensitivity 100, 20, or 2 asexual parasites per µL. Simulation of each EIR, coverage, and MSAT diagnostic sensitivity was repeated for 100 stochastic realizations.

Elimination was determined to be achieved if true asexual parasite prevalence was 0 for the last 150 days of simulation year 3. Probability of elimination was the fraction of the 100 stochastic realizations resulting in elimination.

**Results**

*Composition of the infectious reservoir in the absence of interventions*

Adaptive immunity and human infectiousness parameters within an agent-based model were calibrated to age- and season-stratified data from two high-transmission areas in Burkina Faso (**Figures 1 and 2**) (see Methods). Typical infection trajectories in the calibrated model show shorter durations of infection and lower parasite densities in adults, especially in high-transmission settings (**Figure 1B** and **Figure S4 in Additional file 2**). Gametocyte clearance lagged behind asexual parasite clearance.

**Figure 3A** shows the infectious reservoir over 3 years in a moderate-transmission setting with annual EIR of 10 and the seasonality of Burkina Faso. Total asexual parasite prevalence peaks toward the end of the rainy season and is largely due to a large increase in prevalence of RDT-positive infections. Prevalence of RDT-negative, microscopy positive infections as well as infections positive by PCR only and undetectable infectious remained relatively more constant through the year, peaking toward the beginning of the dry season as the RDT-positive infections begin the clear.

Total infectiousness was quantified by measuring the probability of each individual infecting a mosquito and scaling by age to approximate the decreased biting risk of children due to smaller surface area (see Methods). Total infectiousness exhibits more seasonal variation than asexual parasite prevalence since transmission to mosquitoes is dominated by RDT-positive individuals, and higher parasite density leads to greater infectiousness.

Under these conditions of moderate transmission and high seasonality, the composition of the infectious reservoir shows considerable variation through the year. RDT-positive individuals comprise over 60% of the infectious reservoir during the wet season but only 30% during the dry season. Submicroscopic infections account for 20-40% of the infectious reservoir at any time of year, a substantial portion.

While children have higher parasite densities and are highly infectious (**Figure 1 and 2**), their smaller surface area reduces their contribution to the total infectiousness of the population. Children contribute the most to the infectious reservoir at the beginning of the dry season when adult immune systems have already reduced parasite load in adults, leaving children to slowly clear their longer infections and continue transmission (**Figure 1B**).

To compare the infectious reservoir across a range of transmission intensities, we averaged the seasonal variation in asexual parasite prevalence and total infectiousness over a year of simulation (**Figure 3B**). Total asexual parasite prevalence increases steeply as EIR increases from < 0.1 to 1 infectious bite per person per year. Above EIR = 1, asexual parasite prevalence increases more gradually and approaches 100% for EIR > 100. Total population infectiousness also increases steeply as EIR increases toward 1 but is nearly constant for EIRs between 1 and 200, consistent with observations in a study area where the proportion of infected mosquitoes remained roughly constant while slide prevalence in humans declined nearly four-fold[33].

RDT-detectable infections comprise 20-30% of all infections over the entire sampled range of transmission intensities, with higher EIR settings showing relatively more RDT-detectable infections (**Figure S5 in Additional file 2**). These observations are consistent with the proposed hypothesis that individuals in high-transmission settings are constantly subjected to bursts of high parasite density from new infections, while individuals in low-transmission settings experience a long period of low-density infection as untreated single infections are cleared by immune activity[12].

RDT-detectable infections represent a similar proportion of the infectious reservoir regardless of transmission intensity, as do microscopy-detectable and PCR-detectable infections. Populations in regions with very low transmission intensity (EIR < 1) have slightly more of the infectious reservoir contributed from RDT-positive individuals, while RDT-negative individuals form an increasingly large portion of the reservoir as EIR increases above 1.

Children contribute more to the infectious reservoir at the highest transmission intensities (EIR > 100). At high transmission intensity, adult immunity retains memory of a large repertoire of antigenic variants, and therefore parasite density is unlikely to remain above the RDT detection limit for long. In this situation, children comprise a larger portion of the RDT-positive population. At low transmission (EIR < 1), children under 5 barely contribute to the infectious reservoir, and children between 5 and 15 contribute only 20%. Adults are the major drivers of transmission in all but the highest transmission settings.

*Composition of the infectious reservoir after recent ramp-up in case management and ITN use*

Regions considering elimination are likely to have already implemented vector control and strengthened their health systems. We considered the effect of good case management and aggressive campaigns with insecticide-treated nets (ITNs) on the infectious reservoir. Both case management and ITN campaigns decreased the observed EIR (**Figure S6 in Additional file 2**).

Under case management with artemether-lumefantrine (AL), where 95% of the population has access to care, 60% of clinical cases seek treatment, and 95% of severe cases seek treatment, the total infectious reservoir is reduced (**Figure 4A**) compared to simulations without case management (**Figure**

**3B**). At low transmission (EIR < 1), case management is particularly effective at depleting the infectious reservoir, as infections are more likely to become clinical cases and treatment of each case may clear a significant portion of the infectious reservoir.

We simulated an ITN campaign where ITN use was skewed toward children (see Methods), as has been recommended for control situations [34]. Deployment of ITNs substantially reduces the infectious reservoir at all transmission levels (**Figure 4B**). Children under 5 comprise an even smaller portion of the infectious reservoir than the case with no ITNs, even at high transmission.

Relative to no interventions, high case management and ITN usage bias the infectious reservoir toward RDT-negative and submicroscopic infections, especially for low-transmission settings. At the same observed annual EIR of 1, settings where EIR has recently been reduced by case management or ITNs show a larger fraction of the infectious reservoir stemming from subpatent infections relative to settings where EIR has historically been at 1 (**Figure 4C**). Settings where EIR has recently been reduced to 1 have populations whose immune systems are adapted to EIR > 1, so quickly reducing EIR leads to infections with lower parasite density. If the reduced EIR is maintained for many years, population immunity will rebound[35], and the composition of the infectious reservoir will lean more toward RDT-positive individuals.

*Depletion of the infectious reservoir after mass drug campaigns*

We tested the probability of elimination following mass screen-and-treat (MSAT) campaigns with dihydroartemisinin-piperaquine (DP) at various levels of diagnostic sensitivity, coverage, and transmission intensity (**Figure 5A**). Three rounds of MSAT with independent coverage were applied during the dry season for two years. An improved pediatric formulation of DP was administered to avoid under-dosing in young children [32, 36]. The MSAT campaign outcomes were compared to mass drug administration (MDA), where all individuals are treated, as well as MSAT campaigns where case management and ITNs have recently reduced EIR.

High coverage, high diagnostic sensitivity, and low transmission all increase the likelihood of elimination following an MSAT campaign. Higher coverage cannot completely compensate for lower diagnostic sensitivity, and elimination is not possible for settings with EIR > 1 when an MSAT is conducted with current RDTs, which have sensitivity around 100 parasites per µL. A field trial of an MSAT campaign in a Burkina Faso site with asexual parasite prevalence between 30 and 50% showed little success with long-term prevalence reduction[37].

Case management and ITN use increase the likelihood of elimination by MSAT. Under the same transmission intensity, lingering population immunity helps push a region toward elimination after a drug campaign if EIR has recently been reduced (**Figure 5B**). The bonus from lingering immunity increases with better coverage, underscoring the critical importance of treating as many people as possible in a drug campaign. Simulation results also suggest that lingering immunity may be most beneficial for elimination efforts at EIR between 1 and 10, where EIR is low enough that elimination is possible but high enough that adult immunity is strong.

**Discussion**

To interrupt the chain of malaria transmission, it is critical to identify who is transmitting. Here we calibrate a mechanistic model of within-host parasite and immune dynamics to field data on gametocyte densities and transmission probabilities. The model is then used to predict the infectious reservoir of malaria by age group and diagnostic threshold over a wide range of transmission intensities.

While high-quality field data is essential for understanding malaria transmission, models are also invaluable for showing that data from diverse geographic, demographic, and intervention conditions can arise from the same underlying mechanisms. In this work we have also found that in our model framework, malariatherapy data on infection duration is incompatible with prevalence measurements from endemic areas, highlighting a likely area for model improvement.

While adaptive immunity toward asexual parasites is fairly well understood, it remains unclear how sexual stage immunity affects gametocyte production, survival within the host, and ability to continue the parasite life cycle within the vector[10, 24, 38-40]. Our model of host immunity did not include any immunity toward gametocytes, and the trend of lower gametocyte density with age is due entirely to lower asexual parasite densities. Comparison of calibrated simulations of gametocyte density to observed distributions of gametocyte densities does not suggest that sexual stage immunity plays a strong role, at least in the high-transmission settings where the data were collected.

Transmission-blocking immunity, where individuals with high gametocyte counts fail to infect a large portion of feeding mosquitoes, has also been proposed. Our calibration suggests that transmission-blocking immunity may exist, as our model systematically failed to replicate data where individuals with gametocyte densities >1000/µL were sometimes observed to infect <5% of mosquitoes. Our model also does not account for cases where individuals with very low or undetectable parasite density have been observed to infect mosquitoes[41]. We thus anticipate that we may be overestimating the contribution to the infectious reservoir of people with high density infections, implying that RDT-negative infections are even more critical to target than our analysis suggests. Measurement uncertainty of molecular methods could also play a role in overestimation of gametocyte densities[42]. In addition, direct skin feeding is known to infect mosquitoes at a higher rate than membrane feeds[43]. Future data collection on transmission by direct skin feeding or concentration of mature gametocytes in the skin will be invaluable for improving models of malaria transmission and understanding the nature of the infectious reservoir.

Our simulations show that children comprise a large portion of the infectious reservoir only at very high transmission intensity, and adults (aged over 15 years in our simulations) are the main drivers of transmission in low to moderate transmission settings. These results are for annual averages, but the infectious reservoir also varies seasonally, and relative contribution from children increases at the end of the wet season when adult infections have largely cleared. The age structure of the infectious reservoir will also change if mobile adults are reimporting infections from another setting where transmission is less seasonal.

Based on warnings that pediatric dosing of DP is insufficient, our simulations increased dosage of DP over current recommended levels. We find an improvement in drug campaign outcomes compared to previous work, which followed current guidelines for dosing, particularly in increasing the probability of elimination at EIR > 10 at moderately high coverage levels[31]. While correct dosing is important for reducing the chance of recrudescence in individual patients, under-dosing is particularly critical to avoid in an elimination scenario, as interrupting transmission cannot occur when a subgroup of the population can continue to harbor and transmit infection or when a subgroup does not receive the benefits of prophylaxis.

Compared to settings where no interventions have perturbed the EIR for a long period, settings where EIR has recently been reduced show a shift in the infectious reservoir toward younger individuals and toward lower density infections. This may suggest that MSAT campaigns should achieve lower success rates in settings with recently reduced EIR because diagnostics will miss a greater fraction of the infectious reservoir. However, our simulations show that MSAT campaigns are actually more successful in settings with recently reduced EIR because strong population immunity more than compensates for ongoing transmission from low-density infections.

Our simulations of MSAT campaigns predict that proper timing of drug campaigns relative to ITN deployment may be critical to harness the power of lingering immunity in order to drive the region toward elimination. However, there is a lack of data to properly calibrate the rate of immunity decay, and our predictions of likely elimination may be optimistic. While gathering such data may prove extremely challenging, we anticipate that thorough understanding of immunity decay is critical for meaningful modeling of elimination scenarios.

**Conclusions**

Composition of the infectious reservoir varies seasonally, with higher density infections forming a larger portion during the high-transmission season. RDT-negative infections make up a substantial portion of the infectious reservoir over a wide range of transmission intensities. The increased infectiousness of children due to higher gametocyte densities is balanced by decreased propensity to be bitten due to smaller surface area. Adults comprise the largest fraction of the infectious reservoir at low to moderate transmission intensities, while children form the largest portion only in very high transmission settings.

Interventions such as case management and ITN use tilt the infectious reservoir toward submicroscopic infections. Mass campaigns with antimalarial drugs are more successful when they reach a larger portion of the infectious reservoir through more sensitive diagnostics or higher coverage. Proper timing of drug campaigns with seasonal variation in transmission intensity as well as recent deployment of ITNs allows lingering population immunity to help drive a region toward elimination.

**Competing interests**

The authors declare that they have no competing interests.

**Authors' contributions**

JG designed the simulations, performed the analyses, and wrote the manuscript. ALO and JG designed the density calibration simulations. KAM designed the likelihood functions for calibration. ALO and BK contributed data from Burkina Faso study sites. JG, PAE, and EAW conceptualized the project. All authors have reviewed and approved the final manuscript.

**Acknowledgements**

The authors thank Bill and Melinda Gates for their active support of this work and their sponsorship through the Global Good Fund.

**6**:582–588.

5. Bousema T, Okell L, Felger I, Drakeley C: **Asymptomatic malaria infections: detectability, transmissibility and public health relevance.** *Nature Reviews Microbiology* 2014, **12**:833–840.

6. Jeffery GM, Eyles DE: **Infectivity to mosquitoes of Plasmodium falciparum as related to gametocyte density and duration of infection**. *Am J Trop Med Hyg* 1955, **4**:781–789.

7. Da DF, Churcher TS, Yerbanga RS, Yaméogo B, Sangaré I, Ouédraogo JB, Sinden RE, Blagborough AM, Cohuet A: **Experimental Parasitology**. *Exp Parasitol* 2015, **149**(C):74–83.

8. Young MD, Hardman NF, Burgess RW, Frohne WC, Sabrosky CW: **The infectivity of native malarias in South Carolina to Anopheles quadrimaculatus**. *Am J Trop Med Hyg* 1948, **1**:303–311.

9. Bousema T, Drakeley C: **Epidemiology and Infectivity of Plasmodium falciparum and Plasmodium vivax Gametocytes in Relation to Malaria Control and Elimination**. *Clinical Microbiology Reviews* 2011, **24**:377–410.

10. Drakeley CJ, Bousema JT, Akim INJ, Teelen K, Roeffen W, Lensen AH, Bolmer M, Eling WM, Sauerwein R: **Transmission-reducing immunity is inversely related to age in Plasmodium falciparum gametocyte carriers**. *Parasite Immunol* 2006, **28**:185–190.

11. Greenwood BM: **Asymptomatic malaria infections--do they matter?** *Parasitol Today (Regul Ed)* 1987, **3**:206–214.

12. Okell LC, Bousema T, Griffin JT, Ouédraogo AL, Ghani AC, Drakeley CJ: **Factors determining the occurrence of submicroscopic malaria infections and their relevance for control.** *Nature Communications* 2012, **3**:1237.
21

**Figure legends**

Figure 1 - Calibration of asexual parasite and gametocyte densities. (A) The distribution of asexual parasite and gametocyte densities by season and age group in reference data (open circles) and post-calibration simulation (filled circles) is shown for two study sites. Circle area is proportional to number of individuals and normalized for each season of each age group. (B) Example infection trajectories of asexual parasites (blue) and gametocytes (green) after calibration for an age 6 child (solid lines) and age 28 adult (dotted lines) with immune histories corresponding to high-transmission (top) and low-transmission (bottom) settings. Each individual was challenged with a single infectious bite on day 0, with no subsequent biting. No drugs were administered.

Figure 2 - Calibration of human infectiousness to mosquitoes. The relationship between gametocyte density and infectiousness by season and age group in reference data (open circles) and post-calibration simulation (filled circles) is shown for two study sites. Circle area is proportional to number of individuals and normalized for each season of each age group.

Figure 3 - Composition of the infectious reservoir in the absence of interventions. (A) The composition of the infectious reservoir by diagnostic threshold and age in a setting with moderate transmission. A region of EIR = 10 was simulated over 3 years with seasonal rainfall and temperature based on Burkina Faso climate. The entire population was tested for asexual parasite prevalence daily for 3 years with diagnostics at 3 levels of sensitivity: 100 parasites/µL (bottom 3 stripes, pink), 10 parasites/µL (3 stripes second from bottom, yellow), and 0.05 parasites/µL (3 stripes third from bottom, blue). Infections undetectable at 0.05 parasites/µL are shown in gray (top 3 stripes). Infectiousness was calculated by simulating a membrane feeding test and subsequently scaling by age to approximate surface area effects. The fraction of the infectious reservoir is the normalization of total infectiousness. The average of 100 stochastic realizations is shown. (B) Annual average composition of the infectious reservoir under a range of transmission intensities. All measurements were averaged over the second year of simulation. 1000 stochastic realizations were averaged for each EIR level.

Figure 4 - Composition of the infectious reservoir by diagnostic threshold and age with case management and deployment of insecticide-treated nets (ITNs) under a range of transmission intensities. (A) The infectious reservoir under a range of transmission intensities when case management is available. All treated individuals received a full course of artemether-lumefantrine (AL) with age-dependent dosing. See Methods for details of treatment availability and conditions for seeking care. (B) The infectious reservoir under a range of transmission intensities after deployment of ITNs. See Methods for details of ITN distribution and effectiveness. Scaling human infectiousness for age biting risk included effects of net usage. (C) Comparison of the infectious reservoir at apparent EIR of 1 under conditions of no intervention, case management with AL, and ITN use.

Figure 5 - Depletion of the infectious reservoir after mass drug campaigns with dihydroartemisinin-piperaquine (DP). (A) Probability of elimination after 2 consecutive years of mass drug campaigns at 3 rounds per year. See Methods for details of timing of campaign rounds. Case management and ITNs were simulated as in Figure 4. Probability of elimination was the fraction of 100 stochastic realizations resulting in complete elimination of all parasites by the end of year 3. Coverage was independent for all rounds and all interventions. For simulations with case management and ITNs, EIR is the EIR that would have been experienced during the second year of simulation had the drug campaigns not been administered. Crosses indicate the EIR, coverage, and diagnostic sensitivity simulated in panel B. (B) The infectious reservoir before, during, and after MSAT campaigns in areas with apparent EIR = 1, diagnostic sensitivity 20/μL, and 80% coverage. Left: asexual parasite prevalence and human infectiousness in an endemic region with EIR 1 and no interventions. Right: asexual parasite prevalence and human infectiousness in a region where case management and ITN campaigns have reduced EIR to 1.

**Additional files**

Additional file 1 – Supplementary Tables

Table S1. Parameter values after calibration to prevalence, incidence, and density data. Table S2. Parameter values after calibration to infectiousness data. Table S3. Apparent EIRs tuned by scaling available larval habitat.

Additional file 2 – Supplementary Figures

Figure S1. Comparison of incidence, prevalence, peak density, and infection duration between reference data and simulation with calibrated immunity and gametocyte development parameters. Figure S2. Comparison between simulation and reference data after calibration of immune and gametocyte parameters with full weighting of malariatherapy data. Figure S3. Sensitivity of the composition of the infectious reservoir to stochastic variation and uncertainty in parameter values for high- and moderate-transmission settings by age group and asexual parasite detection limit. Figure S4. Trajectories of infection for children and adults in high- and low-transmission settings with calibrated immune and gametocyte parameters. Figure S5. Unscaled total infectiousness by diagnostic threshold and age group. Figure S6. Malaria transmission under case management and ITN use.

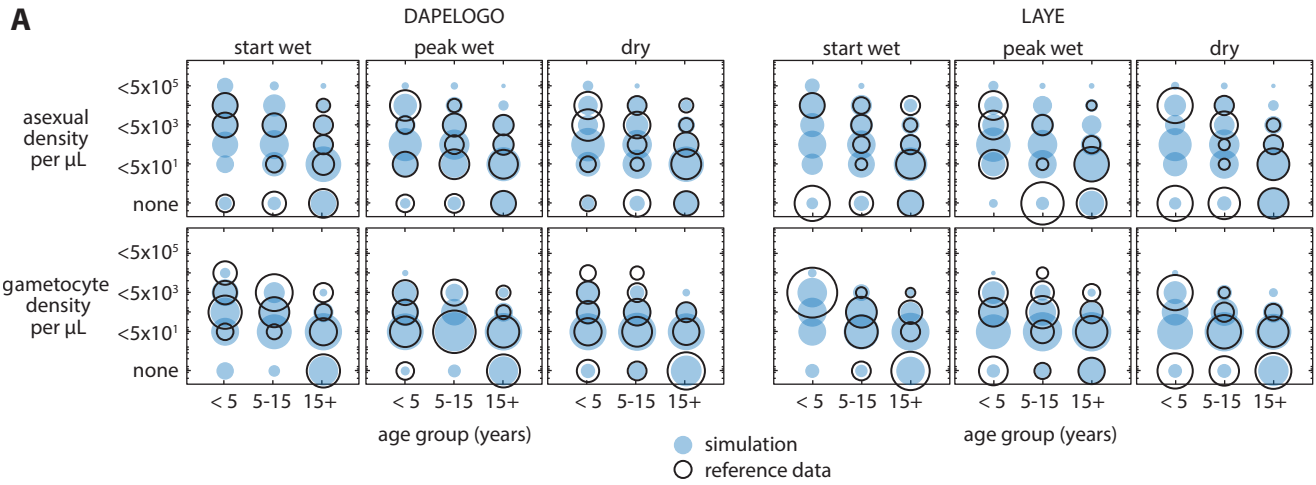
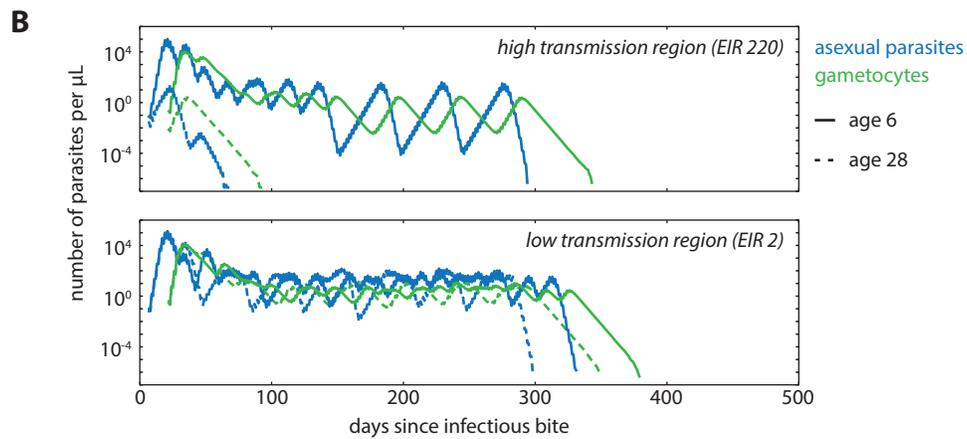

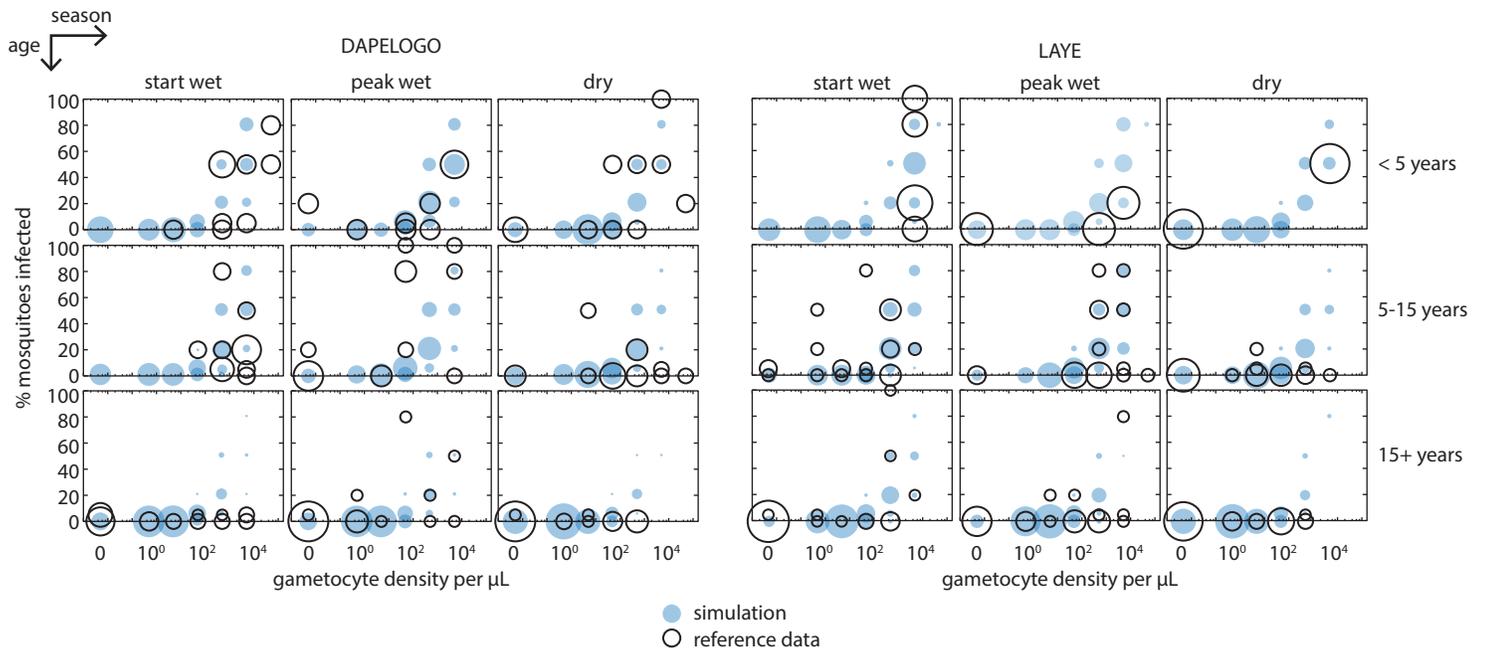

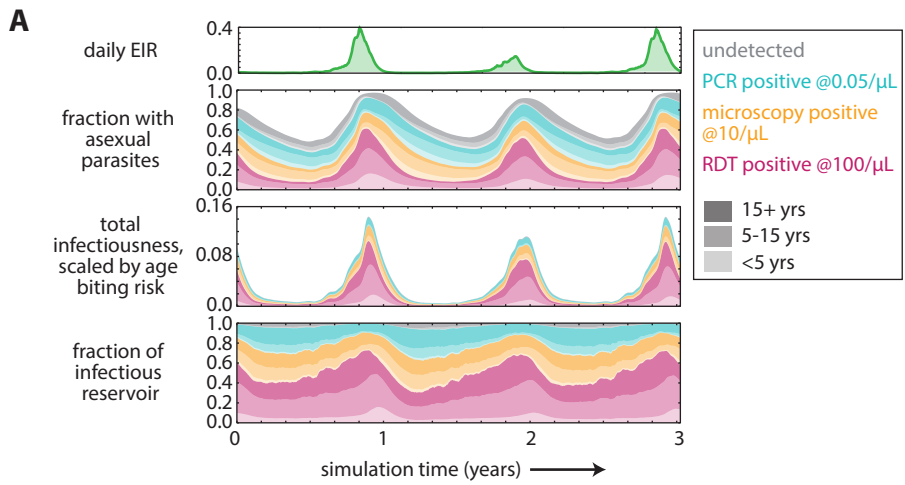
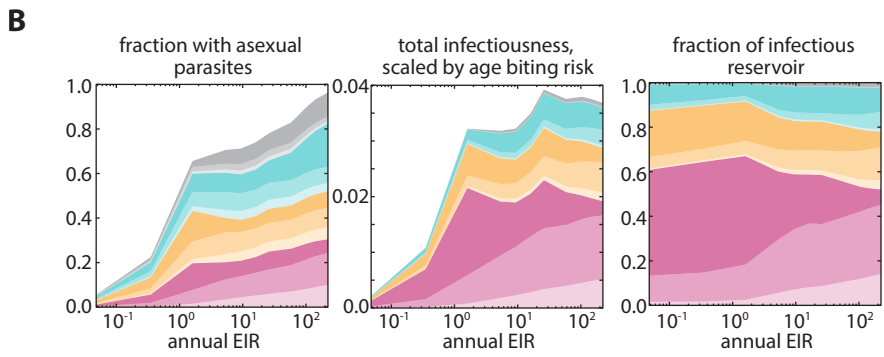

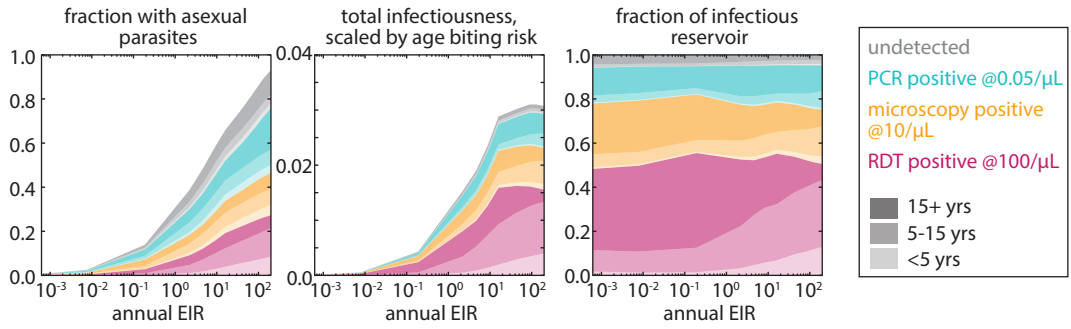
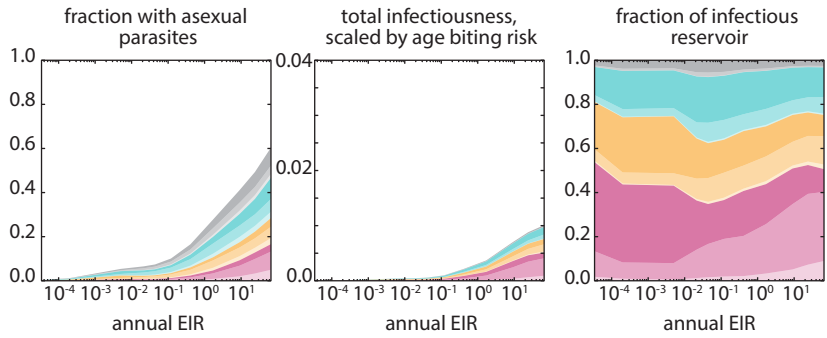
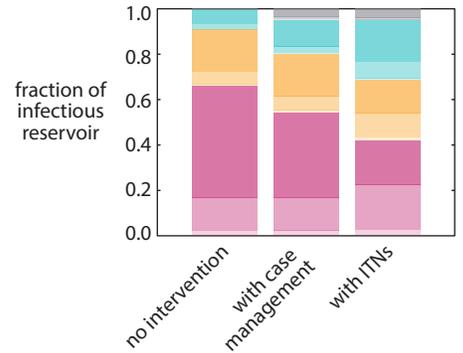

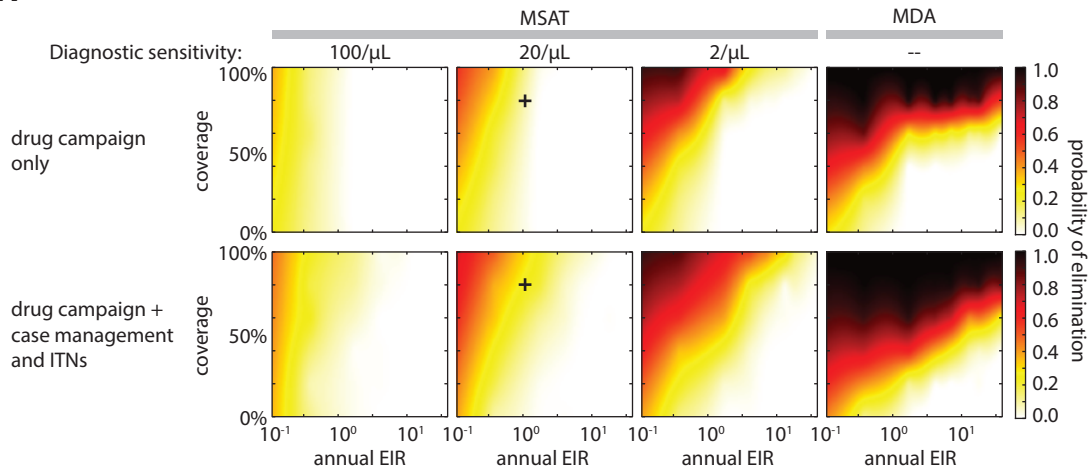

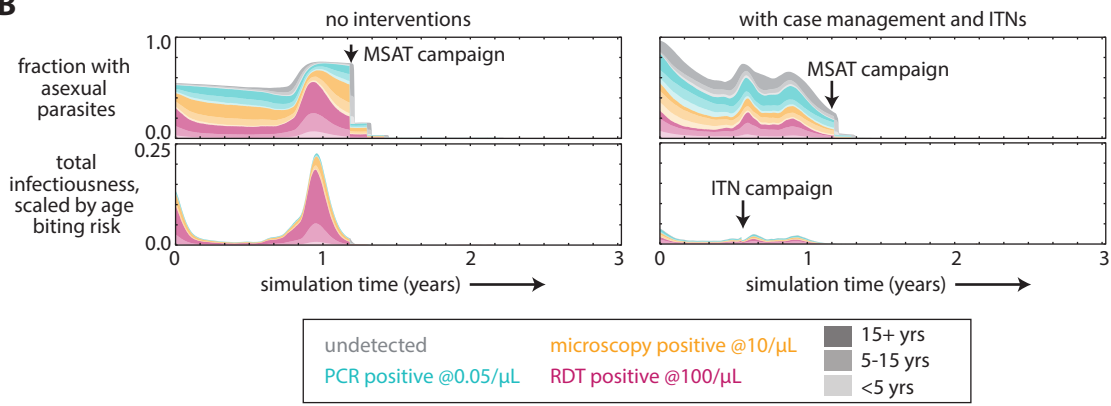

Table S1. Parameter values after calibration to prevalence, incidence, and density data

| Parameter name | Parameter Value |
| --- | --- |
| Max_Individual_Infections | 3 |
| Antigen_Switch_Rate | 2.96e-9 |
| Falciparum_PfEMP1_Variants | 1112 |
| Falciparum_MSP_Variants | 7 |
| MSP1_Merozoite_Kill_Fraction | 0.43 |
| Falciparum_Nonspecific_Types | 90 |
| Nonspecific_Antigenicity_Factor | 0.42 |
| Base_Gametocyte_Production_Rate | 0.044 |
| Gametocyte_Stage_Survival_Rate | 0.82 |

Table S2. Parameter values after calibration to infectiousness data

| Parameter name | Parameter Value |
| --- | --- |
| Base_Gametocyte_Mosquito_Survival_Rate | 0.00088 |
| Acquire_Modifier (vector species parameter) | 0.8 |

Table S3. Apparent EIRs tuned by scaling available larval habitat

| Larval habitat multipler | EIR under no interventions | EIR under case management | EIR with ITNs | EIR with case management and ITNs |
| --- | --- | --- | --- | --- |
| 1.0 | 221 | 201 | 64 | 37 |
| 0.6 | 145 | 128 | 24 | 9.7 |
| 0.4 | 103 | 88 | 8.7 | 3.4 |
| 0.2 | 58 | 44 | 1.5 | 0.57 |
| 0.1 | 27 | 16 | 0.42 | 0.072 |
| 0.08 | 16 | 9.0 | 0.11 | 0.0095 |
| 0.06 | 9.3 | 4.2 | 0.04 | 0.0051 |
| 0.04 | 5.5 | 2.1 | 0.02 | 0.0015 |
| 0.015 | 1.7 | 0.17 | 0.002 | 0.00031 |
| 0.010 | 0.38 | 0.008 | 0.0001 | 0.00005 |
| 0.008 | 0.05 | 0.0003 | 0.00005 | 0.00001 |

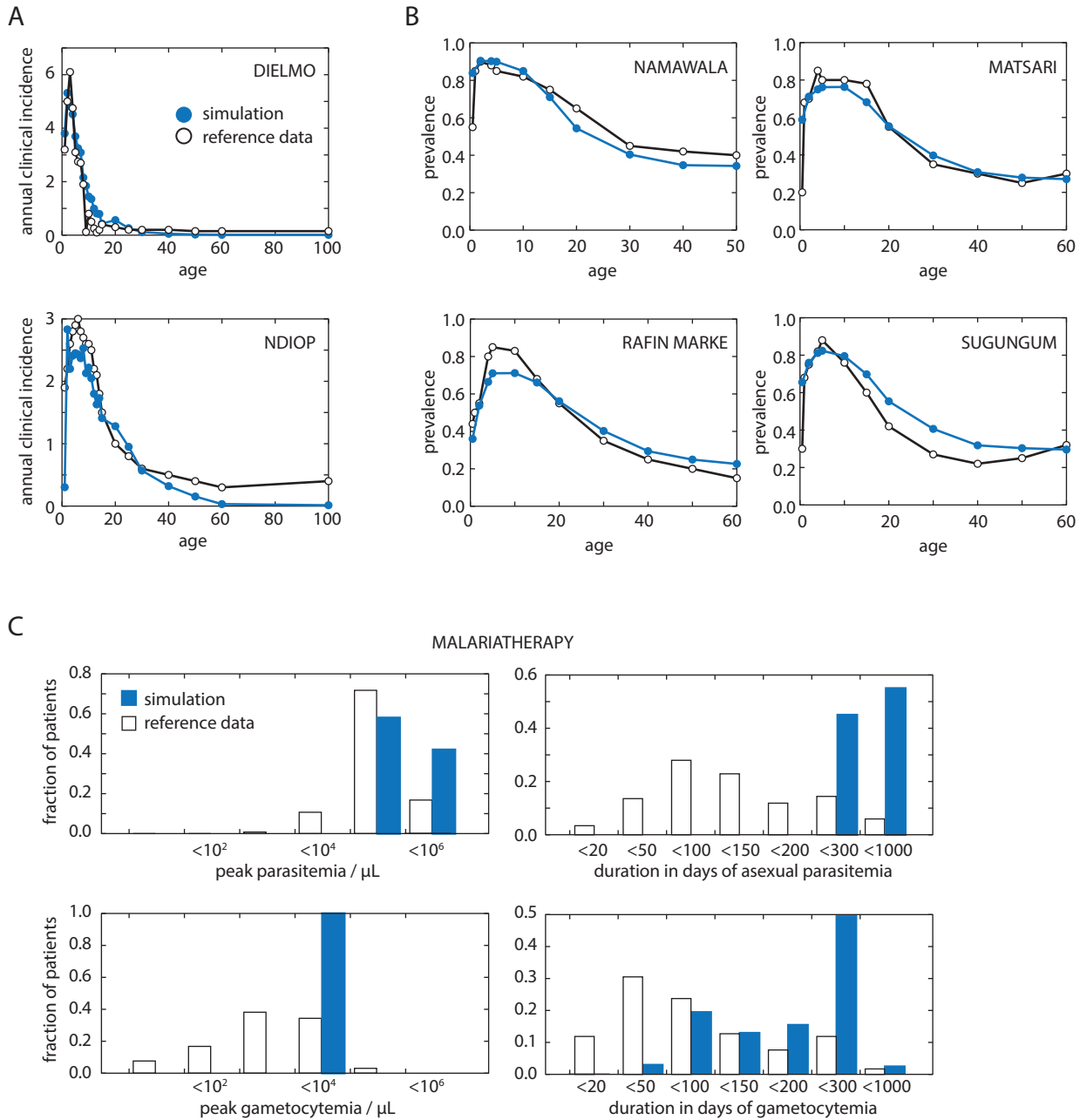

Figure S1. Comparison of incidence, prevalence, peak density, and infection duration between reference data and simulation with calibrated immunity and gametocyte development parameters. Simulation data corresponds to the parameter set with highest combined likelihood across all study sites.

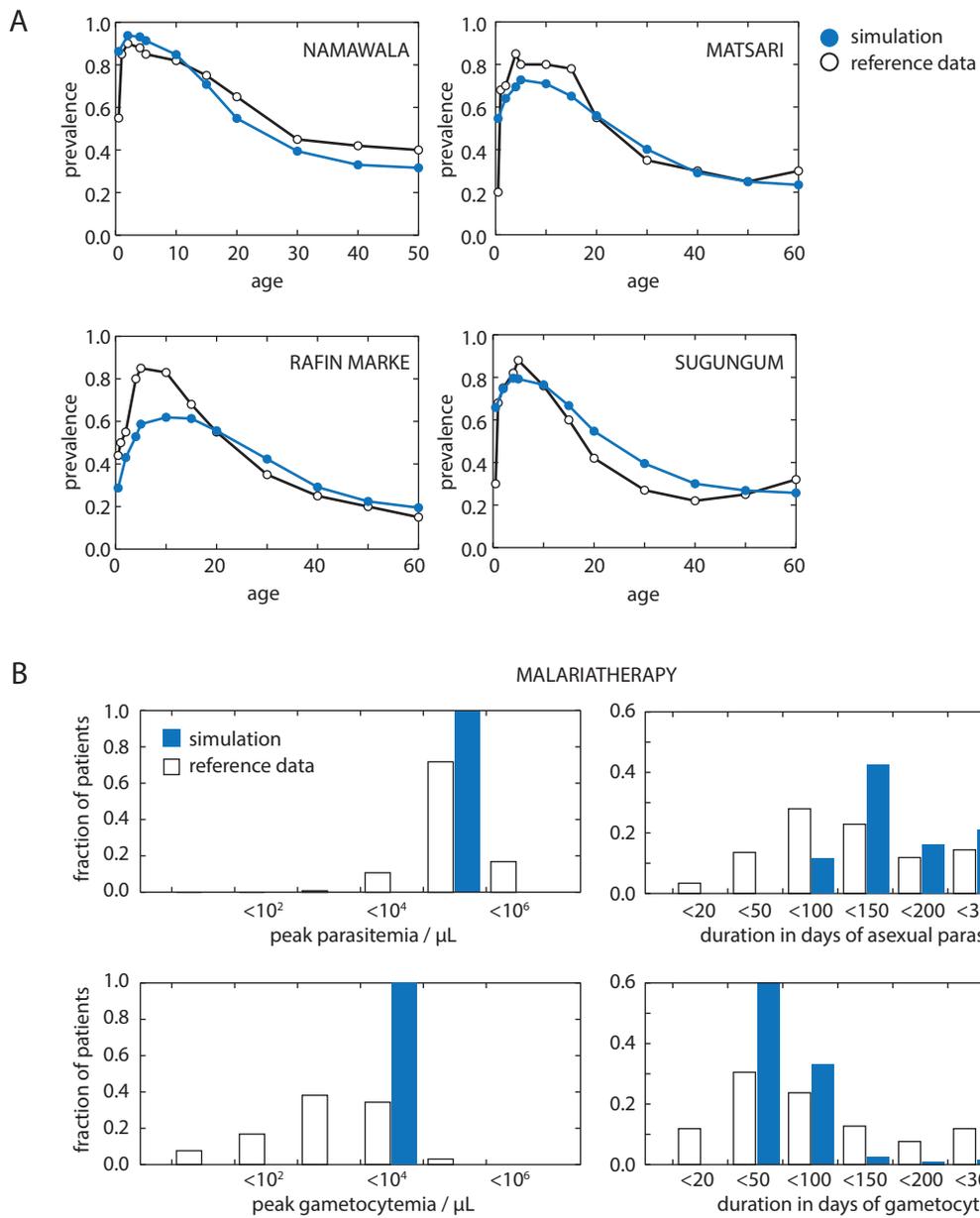

Figure S2. Comparison between simulation and reference data after calibration of immune and gametocyte parameters with full weighting of malariatherapy data. Fit to prevalence data is poorer than in Figure S1.

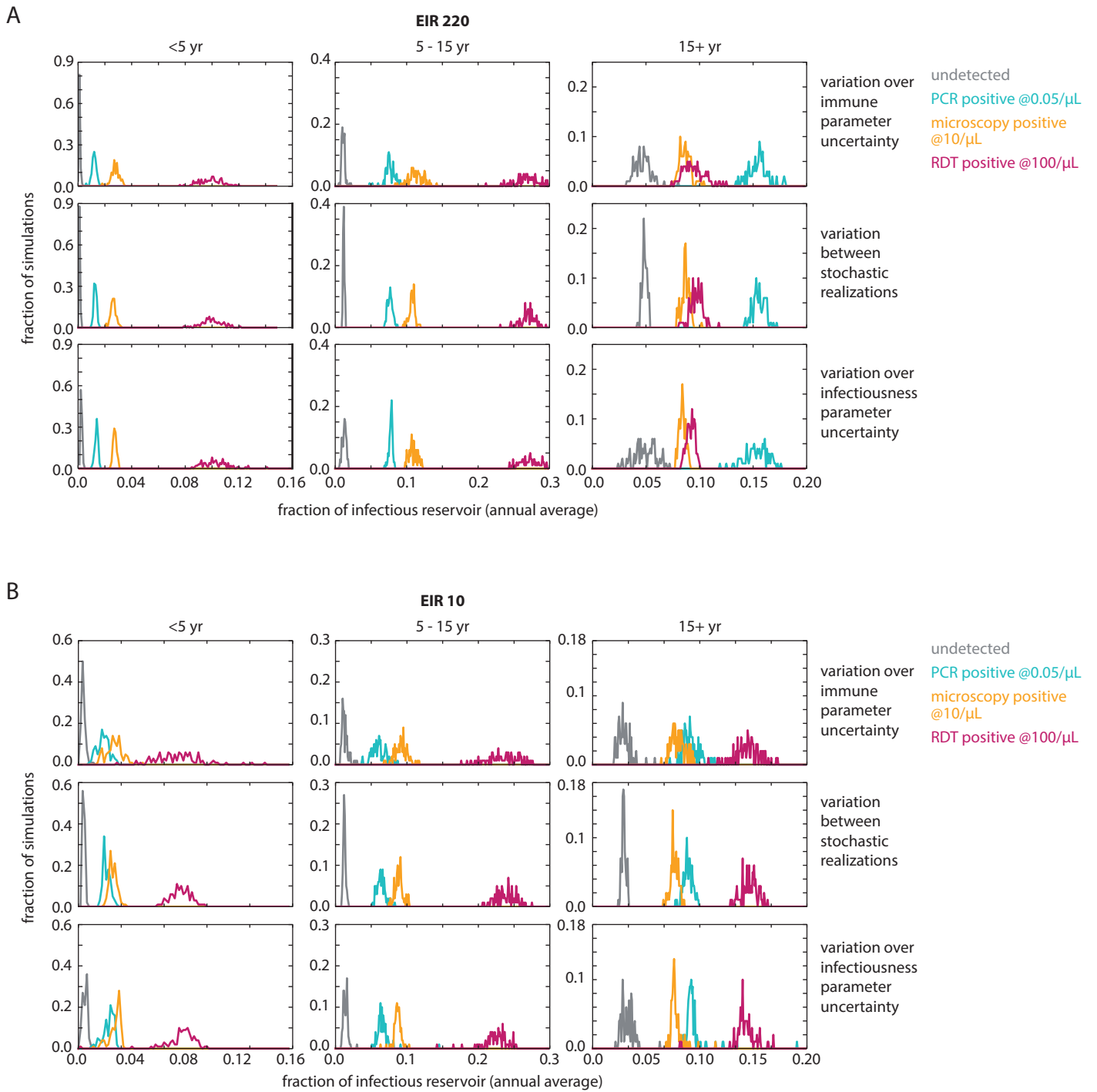

Figure S3. Sensitivity of the composition of the infectious reservoir to stochastic variation and uncertainty in parameter values for (A) high and (B) moderate transmission settings by age group and asexual parasite detection limit. Stochastic variation: 100 realizations of parameters used in Figures 1-5. Parameter uncertainties: 100 highest likelihood parameters from calibrations, 1 stochastic realization per parameter set.

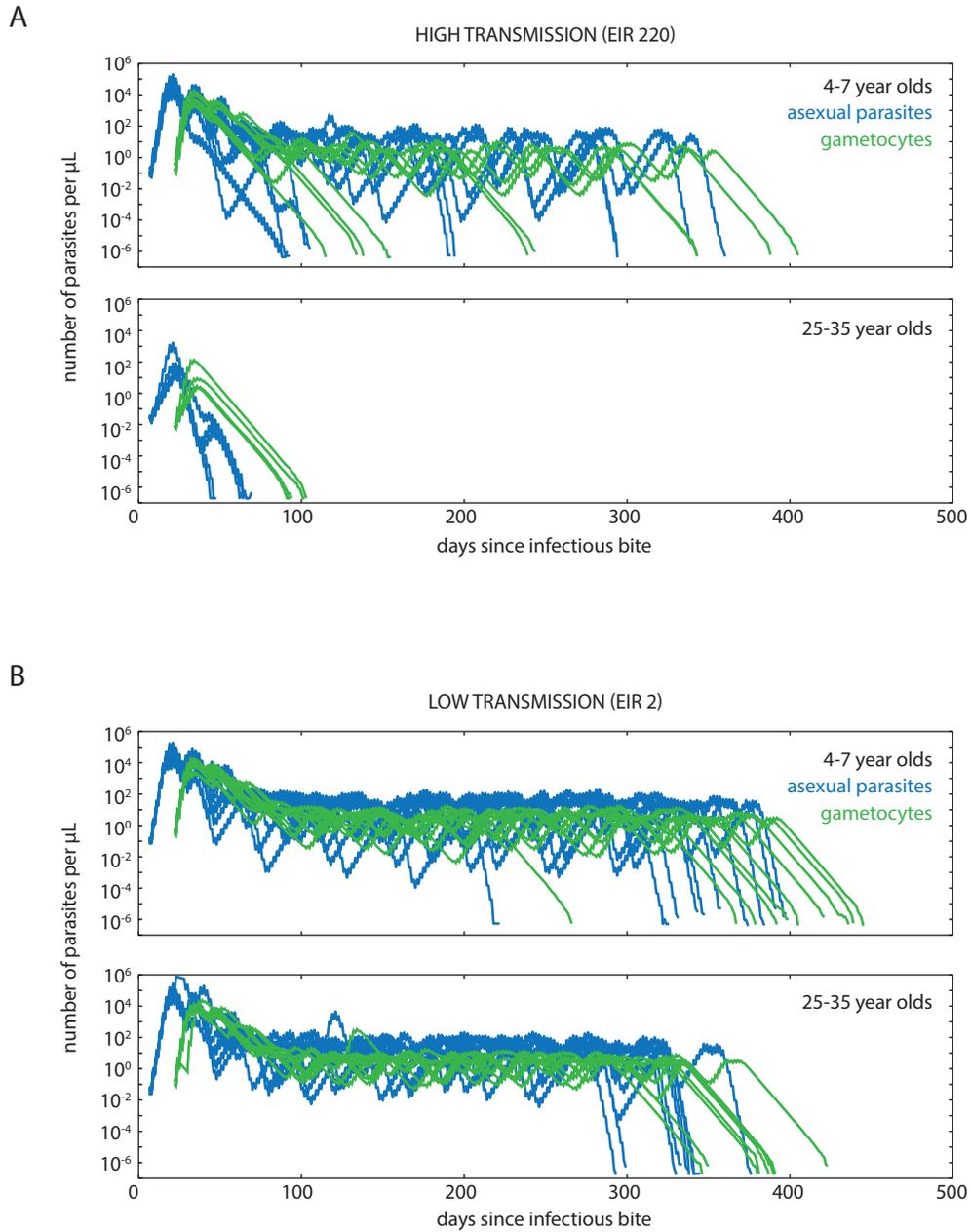

Figure S4. Trajectories of infection for children and adults in (A) high and (B) low transmission settings with calibrated immune and gametocyte parameters. 10 trajectories are shown for each age group. Each individual was challenged with an infectious bite on day 0 without the possibility of additional later infections.

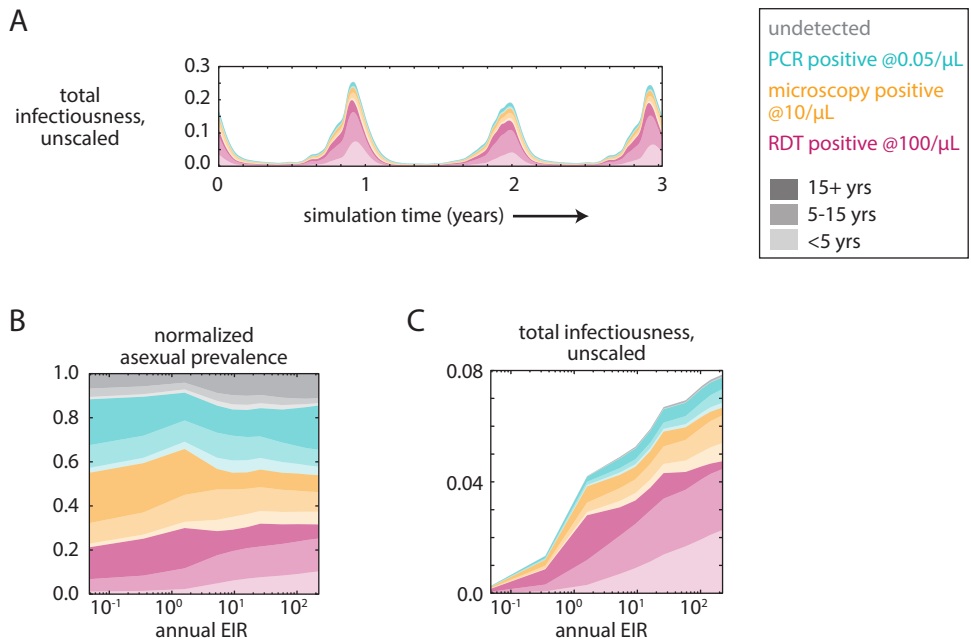

Figure S5. Unscaled total infectiousness by diagnostic threshold and age group. The unscaled total infectiousness is the fraction of mosquitoes that would be infected if the same number of mosquitoes fed on each person in the population. (A) Unscaled total infectiousness over 3 simulation years for setting with EIR = 10. (B) Normalized annual average asexual parasite prevalence over a range of EIRs. (C) Annual average total infectiousness over a range of EIRs.

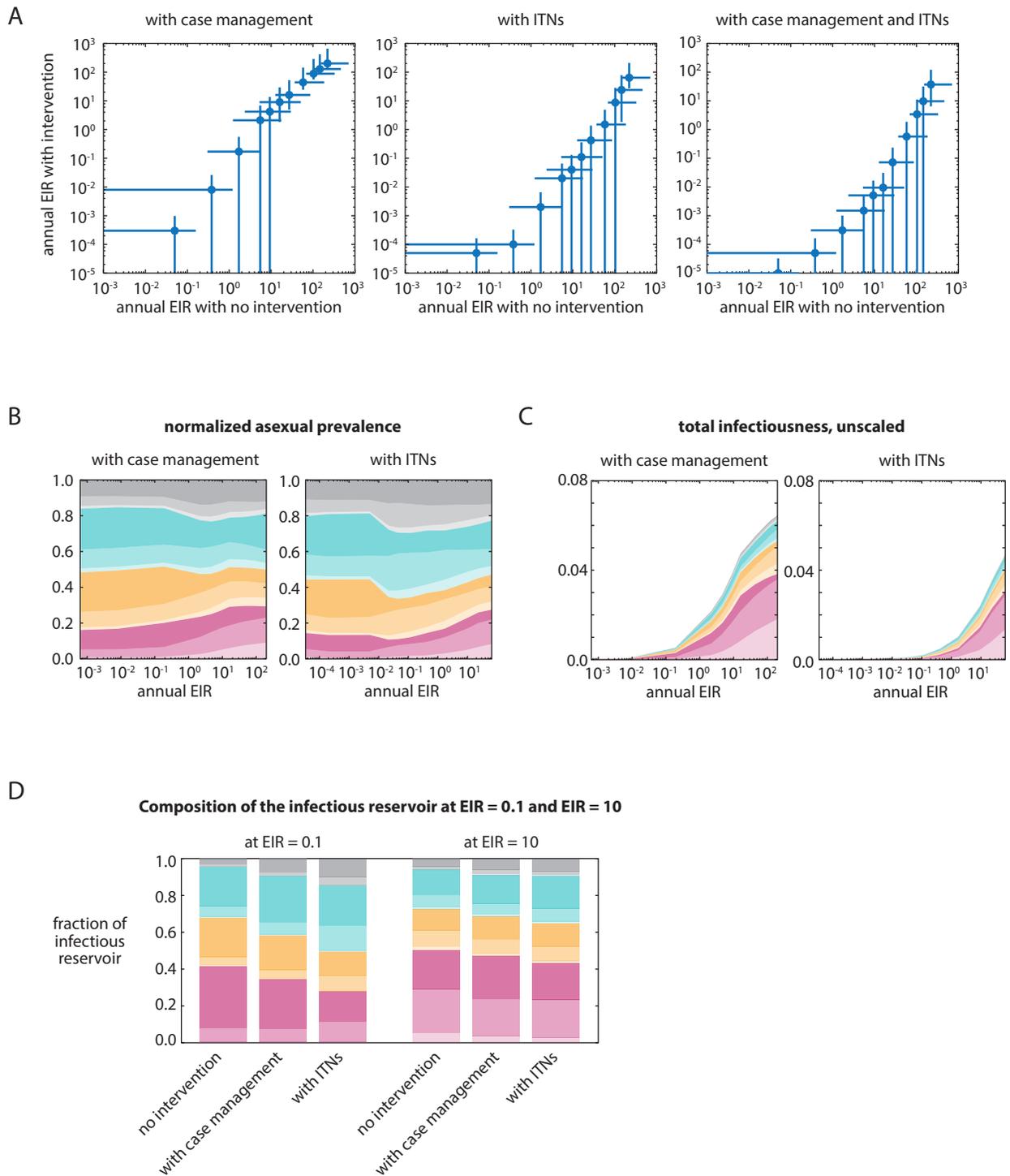

Figure S6. Malaria transmission under case management and ITN use.
(A) Apparent annual EIR drops with case management and ITN campaigns.
(B) Normalized annual average asexual parasite prevalence under case management and ITN use.
(C) Annual average unscaled total infectiousness under case management and ITN use.
(D) Comparison of the infectious reservoir at apparent EIR of 0.1 and 10 for settings with no interventions, case management alone, and ITN use alone.